\documentclass[journal]{igtesymp}
\ifCLASSINFOpdf
\else
\fi
\usepackage[tight,footnotesize]{subfigure}

\usepackage{graphicx}
\usepackage{amsmath}
\usepackage{mathabx}

\begin{document}
%
\title{Towards Real-Time Magnetic Dosimetry Simulations for Inductive Charging Systems}
%
%
%
\author{Norman Haussmann\IEEEauthorrefmark{1},  Martin Zang\IEEEauthorrefmark{1}, Robin Mease\IEEEauthorrefmark{1},  Markus Clemens\IEEEauthorrefmark{1}, Benedikt Schmuelling\IEEEauthorrefmark{3}
and Matthias Bolten\IEEEauthorrefmark{2} 
 \\ \vspace{0.3cm} \normalsize{
\IEEEauthorblockA{\IEEEauthorrefmark{1}University of Wuppertal, Chair of Electromagnetic Theory,\\
 \IEEEauthorrefmark{2}University of Wuppertal, Chair of Scientific Computing and High Performance Computing,\\
  \IEEEauthorrefmark{3}University of Wuppertal, Chair of Electric Mobility and Energy Storage Systems, 42119 Wuppertal, Germany\\
 E-mail: haussmann@uni-wuppertal.de}}
}

%
%


\IEEEaftertitletext{\vspace{-1cm}\noindent\begin{abstract} The exposure of a human by magneto-quasistatic fields from wireless charging systems is to be determined from magnetic field measurements in near real-time. This requires a fast linear equations solver for the discrete Poisson system of the Co-Simulation Scalar Potential Finite Difference (Co-Sim. SPFD) scheme. Here, the use of the AmgX library on NVIDIA GPUs is presented for this task. It enables solving the equation system resulting from an ICNIRP recommended human voxel model resolution of 2\;mm  in less than 0.5\;seconds on a single NVIDIA Tesla V100 GPU.
\end{abstract}
\noindent\vspace{\baselineskip}}

\maketitle
\thispagestyle{empty}\pagestyle{empty}

%
\IEEEpeerreviewmaketitle

\begin{figure}[b]
	\centerline{\includegraphics[width=7.2cm]{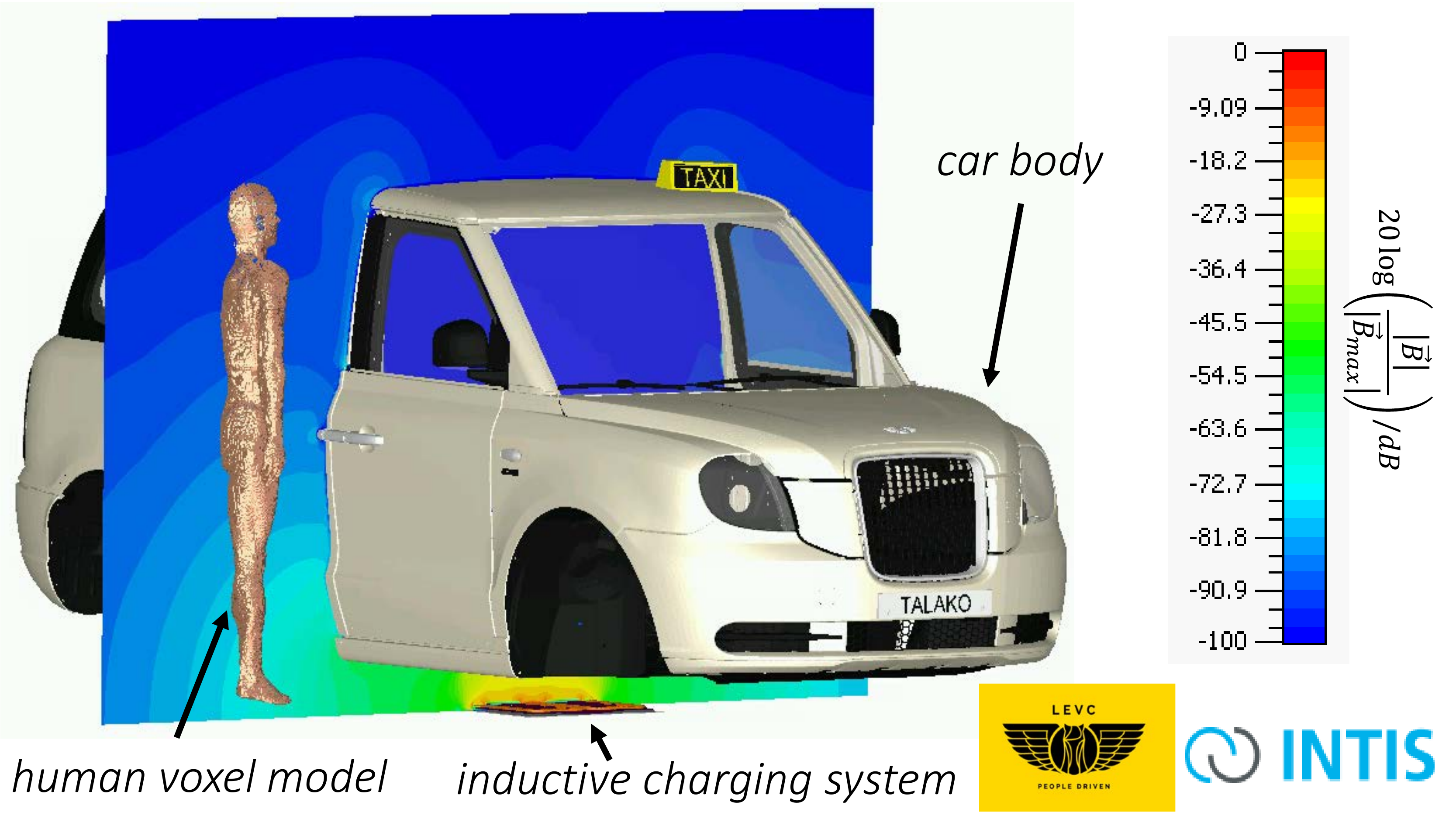}}
	\caption{Exemplary setup of the TALAKO project. The magnetic flux density distribution caused by an inductive charging system (company INTIS  \cite{INTIS}) including the car (company LEVC \cite{LEVC}) body, shielding the human from the magnetic field. The human voxel model is ``Duke'' from ``The virtual family'' \cite{Christ:2009} with a $2\times2\times2\;\mathrm{mm^3}$ resolution.}
	\label{fig_overview}
\end{figure}

\section{Introduction}

Wireless charging systems are a topic of contemporary research for a wide range of applications. They are used e.g. to charge batteries of electrically powered cars. To provide short charging times, the transferred power needs to be in the order of tens of kilowatts. 

Human beings and other biological organisms close to these wireless charging systems are exposed to magneto-quasistatic (MQS) fields excited by low frequency source currents. These MQS fields are potentially harmful to the human body due to induced electric fields which are thus subject to limitations. Since it is difficult or even impossible to perform in vivo measurements, these exposure scenarios need to be determined by numerical simulations.

The International Commission on Non-Ionizing Radiation Protection (ICNIRP) \cite{ICNIRP:2010} recommends a  voxel resolution of $2\times 2\times 2 \;\mathrm{mm^3}$ of the human body for these simulations. Utilizing the Co-Simulation Scalar Potential Finite Difference (Co-Sim. SPFD) \cite{Zang:2017} scheme this is equivalent to $8.9$ million degrees of freedom (DOFs) for a Poisson system of linear equations and the used human voxel model that is named ``Duke'' from ``The Virtual Family'' \cite{Christ:2009}.

The overall goal of this work is to be able to calculate the electric fields inside the human body induced by the MQS field in less than $5\;\mathrm{s}$, what we call ``almost real-time'', starting from magnetic flux density values measured in situ or simulated. Consequently, the electric field averaged over each voxel is determined and therefrom the potential harm to the human body can be evaluated almost in real-time.

One essential brick to achieve this goal is the possibility to solve the Poisson system of linear equations of the Co-Sim. SPFD method as quickly as possible. The preferred solution utilizing the performance of GPUs is presented in this paper.

\section{Approach}
The TALAKO project \cite{Talako:2020} works on a concept to charge batteries of taxis via an inductive charging system. An examplary configuration and the resulting low frequency magnetic field is visualized in Figure \ref{fig_overview}. The human can stand next to the car, sit in the car or even be lying next to the car. In all scenarios the exposure by the MQS field has to be determined. This configuration serves as an exemplary blueprint.

Instead of a pure monolithic- or co-simulation with commercial tools, the idea is to measure the magnetic flux density in situ. The human body has a negligible attenuating effect onto the source field and therefore the exposure can be calculated in a two-step procedure from the measured values. The goal is to get an almost immediate result on the potential harmfulness of the MQS field exposure of the human body by comparing the calculated and voxel averaged electric field values with those given by the ICNIRP basic restrictions. 

As the human model consists of voxels, the MQS exposure simulation method of choice is the SPFD scheme \cite{Dawson:1996a}, \cite{Dawson:1996b}, which has been formulated in terms of the Finite Integration Technique (FIT) \cite{Weiland:1996} in \cite{Barchanski:2005}. To use any MQS/eddy current solver, the Co-Simulation SPFD scheme is used \cite{Zang:2017}. 

Within the SPFD scheme the electric field strength $|\vec{E}|$ at each node of the cartesian grid is calculated from the electric grid voltages $\wideparen{\underline{\boldsymbol{e}}}$

\begin{equation}
\wideparen{\underline{\boldsymbol{e}}} = -j\omega \left [ \wideparen{\underline{\boldsymbol{a}}} + \boldsymbol{G} \underline{\boldsymbol{\Psi}}\right ],
\label{e_field}
\end{equation}
with the discrete magnetic vector potentials $\wideparen{\underline{\boldsymbol{a}}}$, the discrete gradient operator $\boldsymbol{G}$, the discrete vector of time integrals of the electric nodal potentials  $\underline{\boldsymbol{\Psi}}$ and the angular frequency of the field $\omega = 2\pi f$.

None of the required discrete quantities can be directly extracted from the measured magnetic flux density. The discrete magnetic fluxes in the vector $\wideparen{\wideparen{\underline{\boldsymbol{b}}}}$, which are the magnetic flux densities integrated over each voxel facet, are connected to the required magnetic vector potentials via the curl operator $\boldsymbol{C}$:

\begin{equation}
\wideparen{\wideparen{\underline{\boldsymbol{b}}}} = \boldsymbol{C} \wideparen{\underline{\boldsymbol{a}}}.
\label{bCa}
\end{equation}
In the FIT notation this is an over-determined system of equations that can be disentangled with the tree-cotree gauging technique \cite{Albanese:1990} as implemented within the Co-Sim. SPFD scheme \cite{Zang:2017}.

As the displacement currents can be ignored in MQS conditions, the discrete vector of time integrals of electric nodal potentials $\underline{\boldsymbol{\Psi}}$ can also be determined from discrete magnetic vector potentials. The equation reads as
\begin{equation}
\boldsymbol{G^\mathsf{T}}\boldsymbol{M_\kappa} \boldsymbol{G} \underline{\boldsymbol{\Psi}} = -\boldsymbol{G^\mathsf{T}}\boldsymbol{M_\kappa} \wideparen{\underline{\boldsymbol{a}}},
\label{poisson_eqs}
\end{equation}
which is called the discrete Poisson system of linear algebraic equations of the SPFD scheme. To solve this system of equations, the matrix of electrical conductivities $\boldsymbol{M_\kappa}$ of the human body tissues at the field's frequency needs to be determined.

Although this system of linear equations has reduced DOFs compared to equation (\ref{e_field}), it still consists of approximately $8.9$ million DOFs considering the ICNIRP recommended voxel resolution of $2\times2\times2\;\mathrm{mm^3}$ and the used model ``Duke''. 

After the discrete vector of time integrals of electric nodal potentials is determined from equation (\ref{poisson_eqs}), it can be inserted into equation (\ref{e_field}) alongside the magnetic vector potentials and solved for obtaining the electric grid voltages.

\subsection{Technical Summary}
The steps to determine the electric field inside the human body tissues are summarized in this section.
\begin{enumerate}
	\item Coarse sampling in situ measurements of the magnetic flux density.
	\item Interpolation/extrapolation of the magnetic flux density to determine the magnetic flux on a fine grid of $2\times2\times2\;\mathrm{mm^3}$.
	\item Solve equation (\ref{bCa}) to determine magnetic vector potentials using tree-cotree gauging.
	\item Solve Poisson system of linear equations (\ref{poisson_eqs}).
	\item Calculation of electric grid voltages, equation (\ref{e_field}), and determination of the electric field strength at each node.
	\item Calculation of $99^{\mathrm{th}}$ percentile of the voxel averaged electric field strength and create pristine plots with the electric field of each human tissue voxel.
\end{enumerate}
The goal is to have almost real-time magnetic dosimetry simulations for inductive charging systems, which implies that steps $2)$ to $6)$ are conducted quickly. To have a reasonably fast feedback on the induced electric fields inside the human body tissues, these fields need to be calculated as fast as possible, aiming at simulation times of less than $5\;\mathrm{s}$. 

This paper focuses on step $4)$, which needs to be processed in a fraction of the imposed $5\;\mathrm{s}$ time limit. Hence, the Poisson system of linear equations needs to be solved as fast as possible and Graphics Processing Units (GPUs) are predestined for that task due to their performance with massively parallel algorithms. The premises are to solve the linear equations system in double precision with a total reduction of the relative residual norm of smaller than $10^{-12}$.

\subsection{Simulation Setup}
All steps, except the first, described in the previous section can be tested and compared to commercial simulation tools. Here, CST Microwave Studio \cite{CST} serves as a reference implementation.

To enhance the simulation speed, the setup shown in Figure \ref{fig_overview}  is simplified. It consists of the same wireless charging system, but instead of the full car body only a geometrically simplified version of the steel floor is used.

The reference simulation is performed with the Scaled-Frequency Finite Difference Time Domain method \cite{Gandhi:1992} at $5\;\mathrm{MHz}$. Accordingly, also the conductivity of the steel plate and human body tissues are scaled to that frequency \cite{Cimala:2015}, \cite{Zang:2017b}. To compare the electric field strength inside the human body tissues to the result of the SPFD scheme, the electric field of the CST simulations needs to be scaled to the desired frequency of $f=85\;\mathrm{kHz}$ from $f'=5\;\mathrm{MHz}$  via
\begin{equation}
\vec{E}_\mathrm{body}(f) = \frac{f}{f'} \cdot \frac{\kappa(f')}{\kappa(f)}\cdot\vec{E}_\mathrm{body}(f'),
\end{equation} 
with the frequency dependent electrical conductivities $\kappa$. The entire CST simulation is performed monolithically.

To set up the Poisson system of linear equations of the SPFD scheme, the magnetic field is extracted from CST Microwave Studio to calculate the magnetic flux density at the required positions of a potential body exposure. Hence, the external interpolation of the magnetic flux density is not yet used to determine the magnetic flux on the fine grid, but rather the interpolation included in the CST implementation. From here, the magnetic vector potentials are determined and the Poisson system of linear equations (\ref{poisson_eqs}) is assembled.

The time required to solve this system of linear equations is discussed in the next section.

\section{Numerical Tests}
To determine the electric field in less than five seconds, the Poisson system of linear equations needs to be solved in a fraction of this time limit. Several GPU based libraries with different solvers and preconditioners have been tested to this end and the best results are obtained with an Algebraic Multigrid preconditioner for an iterative Krylov type Flexible Generalized Minimal Residual (FGMRES) solver utilizing the NVIDIA AmgX library \cite{Naumov:2015}.

The performance has been tested on different generations of NVIDIA GPUs. One compute server system consists of a node with five Tesla K20m cards, one node server with a Tesla K80 (a dual) GPU, and one node server with eight Tesla V100 cards.

\subsection{Tesla K20m}
\begin{figure}[ht]
	\centering
	\centerline{\includegraphics[width=7.2cm]{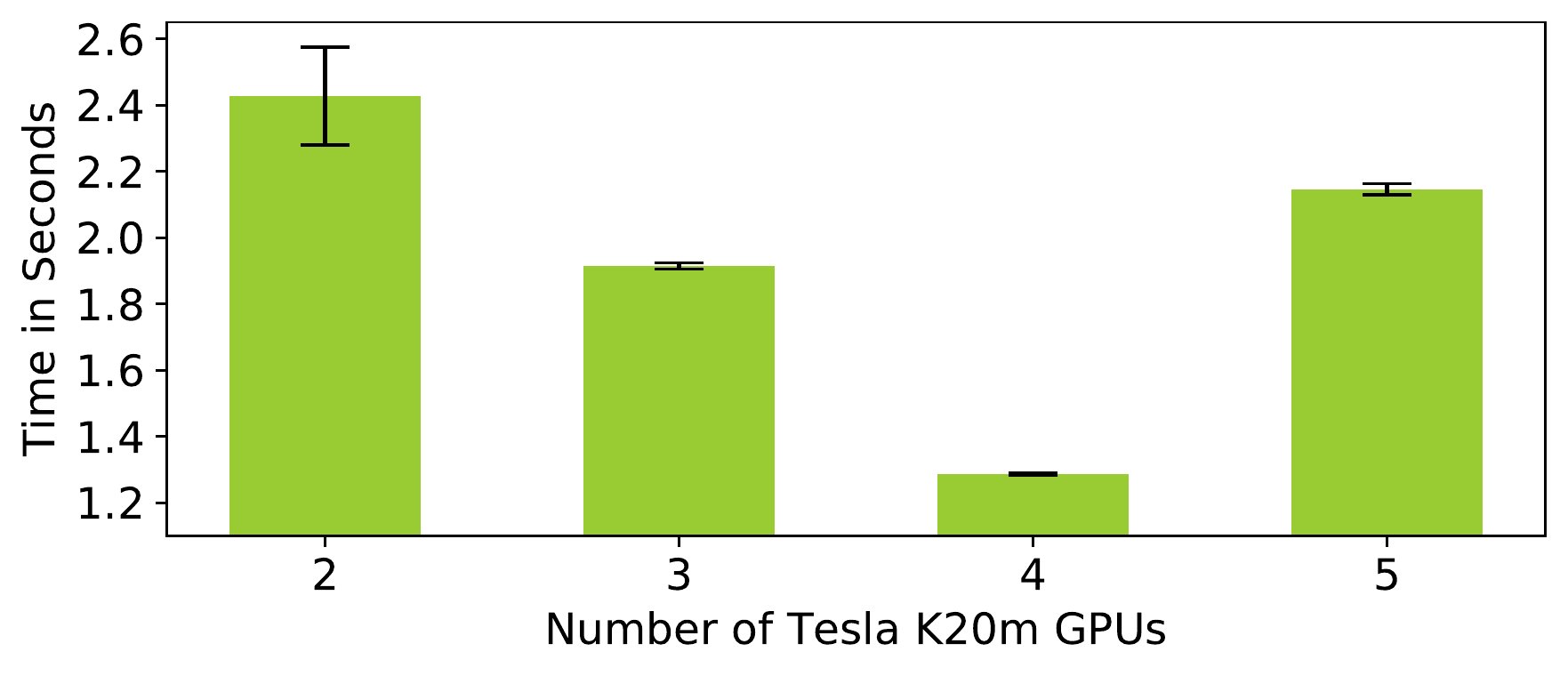}}
	\caption{Required time to solve the system of linear equations of the voxel model ``Duke'' with a $2\times 2\times 2\;\mathrm{mm^3}$ resolution and $8.9$ million DOFs on Tesla K20m.}
	\label{K20m_2mm}
\end{figure}

Using the ICNIRP recommended human voxel model resolution of $2\;\mathrm{mm}$ and the model ``Duke'' leads to $8.9$ million DOFs of the Poisson system of linear equations. Only voxels with electrically conductive body tissues are considered in the model as the electric field strength is only to be simulated inside the body model. In double precision this requires more than $9.5$ GB of memory for the system matrix and the right hand side vector (consisting of real numbers) of the linear algebraic equation system utilizing the preconditioner and solver mentioned above. A Tesla K20m GPU, released in fall 2012, features $5\;\mathrm{GB}$ of memory and $2496$ CUDA cores \cite{NVIDIA:2012}. Hence, at least two GPUs are required in parallel to solve the Poisson system of linear equations. In the given resolution the speed results for a single node with up to five Tesla K20m GPUs is visualized in Figure \ref{K20m_2mm}.  
Each GPU configuration was subjected to five different simulation runs. The mean value of the required CPU/GPU time is shown as columns and its standard deviation as the error bar.

\begin{figure}[ht]
	\centering
	\centerline{\includegraphics[width=7.2cm]{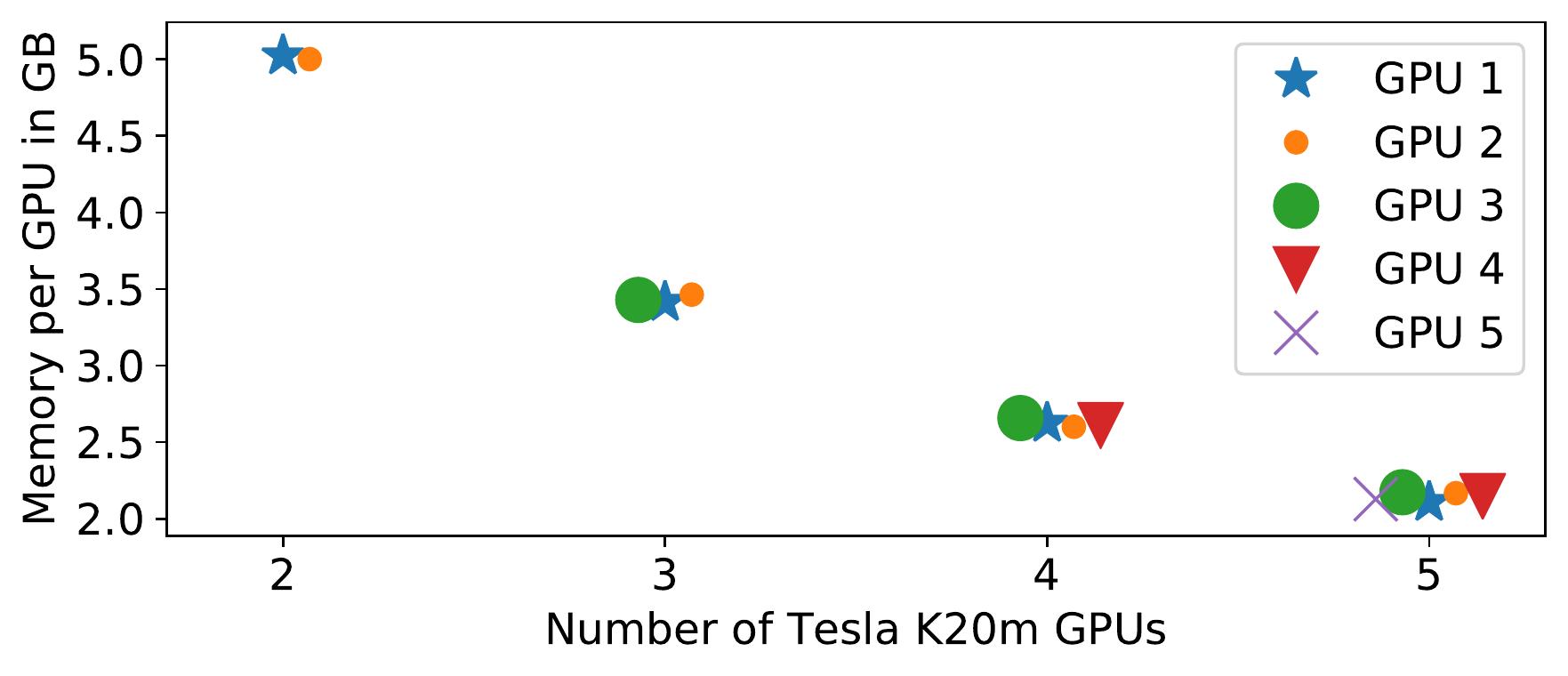}}
	\caption{Maximum required memory on each GPU to solve the system of linear equations of the voxel  model ``Duke'' with a $2\times 2\times 2\;\mathrm{mm^3}$ resolution and $8.9$ million DOFs at different GPU configurations.}
	\label{K20m_2mm_memory}
\end{figure}

Two Tesla K20m cards allow to solve the system of linear equations in less than $2.5\;\mathrm{s}$. The required time to solve the system decreases when increasing the number of GPUs to a maximum of four K20m GPUs. The optimum is found using four GPUs with less than $1.3\;\mathrm{s}$ to solve the equations system. The standard deviation is small with less than $4\;\mathrm{ms}$ and therefore the calculation time is perfectly reproducible. Using all five GPUs available to the node, the required time increases and solving the equations system is even less efficient than using three Tesla K20m.
To check if the linear equations system is distributed uniformly across the GPUs, the required memory is investigated. The maximum required memory per GPU is shown in Figure \ref{K20m_2mm_memory}.

The required memory is equally distributed across the utilized GPUs, but does not decrease linearly with the number of Tesla K20m cards. Looking at the integrated memory usage (see Table \ref{tab_total_mem_usage_K20m_2mm}), the required memory increases with the number of GPUs. This additional overhead increases while the actual data of the linear equation system decreases per GPU. At some number of GPUs, the calculation becomes increasingly inefficient with a growing number of GPUs as the required time to solve the system of linear equations on each GPU is small, but the required intercommunication overhead between the GPUs increases. This needs to be determined for each linear equations system seperately and depends on the DOFs as well as the GPU type. 

\begin{table}
	\centering
	\caption{Integrated memory usage of all utilized Tesla K20m cards to solve the system of linear equations of the voxel  model ``Duke'' with a $2\times 2\times 2\;\mathrm{mm^3}$ resolution.}
	\label{tab_total_mem_usage_K20m_2mm}
	\begin{tabular}{|l|c|c|c|c|}
		\hline
		\textbf{Number of K20 GPUs}& \textbf{2}& \textbf{3} & \textbf{4}& \textbf{5}\\\hline
		\textbf{Memory usage in GB} & $10.026$ & $10.306$ & $10.493$ & $10.723$ \\\hline
	\end{tabular}
\end{table}
\begin{figure}[h]
	\centering
	\centerline{\includegraphics[width=7.2cm]{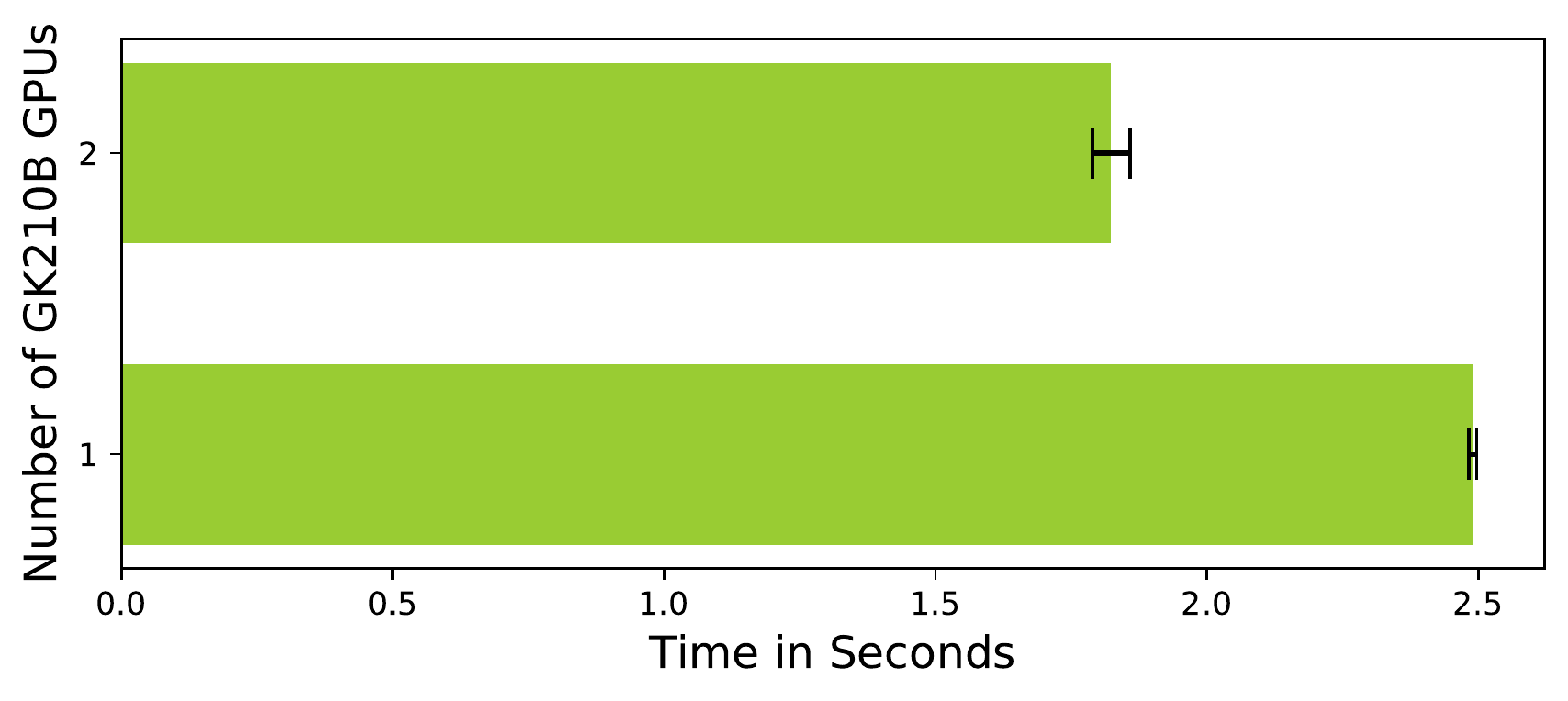}}
	\caption{Required time to solve the Poisson system of the voxel  model ``Duke'' with a $2\times 2\times 2\;\mathrm{mm^3}$ resolution and $8.9$ million DOFs on Tesla K80.}
	\label{k80_2mm}
\end{figure}
\subsection{Tesla K80}
The Telsa K80 was released in fall 2014. It features two Tesla GK210B GPUs with $2496$ CUDA cores and $12\;\mathrm{GB}$ of memory each \cite{NVIDIA:2015}. Thus, a single GPU has sufficient memory to store the Poisson system of linear equations of the ICNIRP recommended voxel model resolution of $2\times 2\times 2\;\mathrm{mm^3}$ ($8.9$ million DOFs) using the model ``Duke''. The required time to solve this equations system can be seen in Figure \ref{k80_2mm}.

A single GK210B GPU has about the same performance as two Tesla K20m GPUs although equipped with the same number of CUDA cores. Utilizing both GPUs of the Telsa K80 improves the time to solve the linear equations system by $37\;\%$ to below $2\;\mathrm{s}$.

\subsection{Tesla V100}

\begin{figure}[ht]
	\centering
	\centerline{\includegraphics[width=7.2cm]{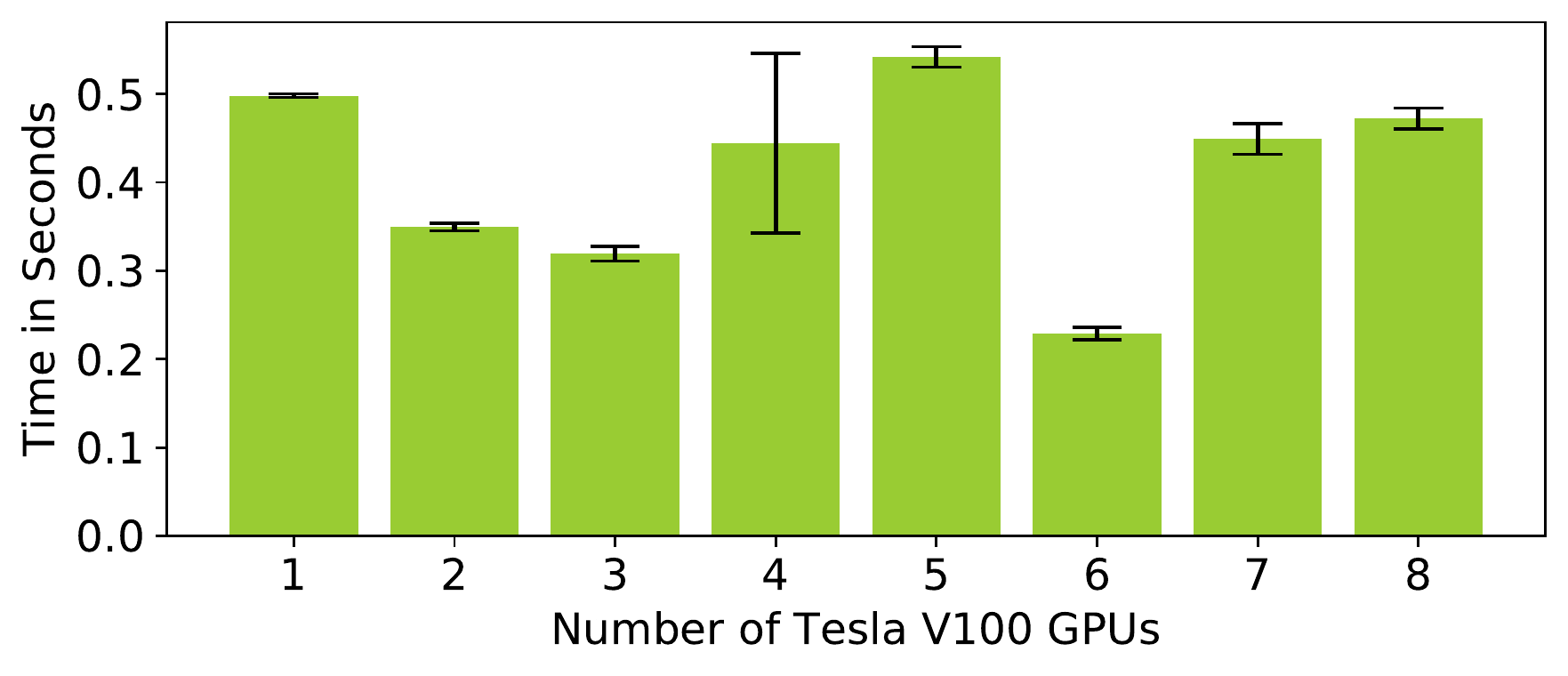}}
	\caption{Required time to solve the Poisson system of linear equations of the voxel  model ``Duke'' with a $2\times 2\times 2\;\mathrm{mm^3}$ resolution and $8.9$ million DOFs on Tesla V100.}
	\label{v100_2mm}
\end{figure}

The NVIDIA Tesla V100 was released in fall 2017 and is therefore half a decade younger than the Tesla K20m and three years younger than the Tesla K80. It is equipped with $32\;\mathrm{GB}$ of memory, $5120$ CUDA Cores and $640$ Tensor Cores \cite{NVIDIA:2020}. Hence, the performance should exceed that of the Tesla K20m and K80 in solving the Poisson system of linear equations with $8.9$ million DOFs. The required time for solving the system of linear equations for the voxel model with a $2\times 2\times 2\;\mathrm{mm^3}$ resolution can be seen in Figure \ref{v100_2mm}.
\\
A single Tesla V100 solves the Poisson system of linear equations with $8.9$ million DOFs in less than half a second with a standard deviation smaller than $2\;\mathrm{ms}$. The solution times show a strong dependence on the number of Tesla V100 GPUs used. An optimum is found with six GPUs as these require less than a quarter of a second ($0.229\;\mathrm{s}$) at a standard deviation of $7\;\mathrm{ms}$ to solve the system of equations. The large differences occur as the number of iteration steps differs up to a factor of two to achieve the desired reduction of the residual.
\subsection{Voxel Model with 1\;mm Resolution}
\begin{figure}[ht]
	\centering
	\centerline{\includegraphics[width=7.2cm]{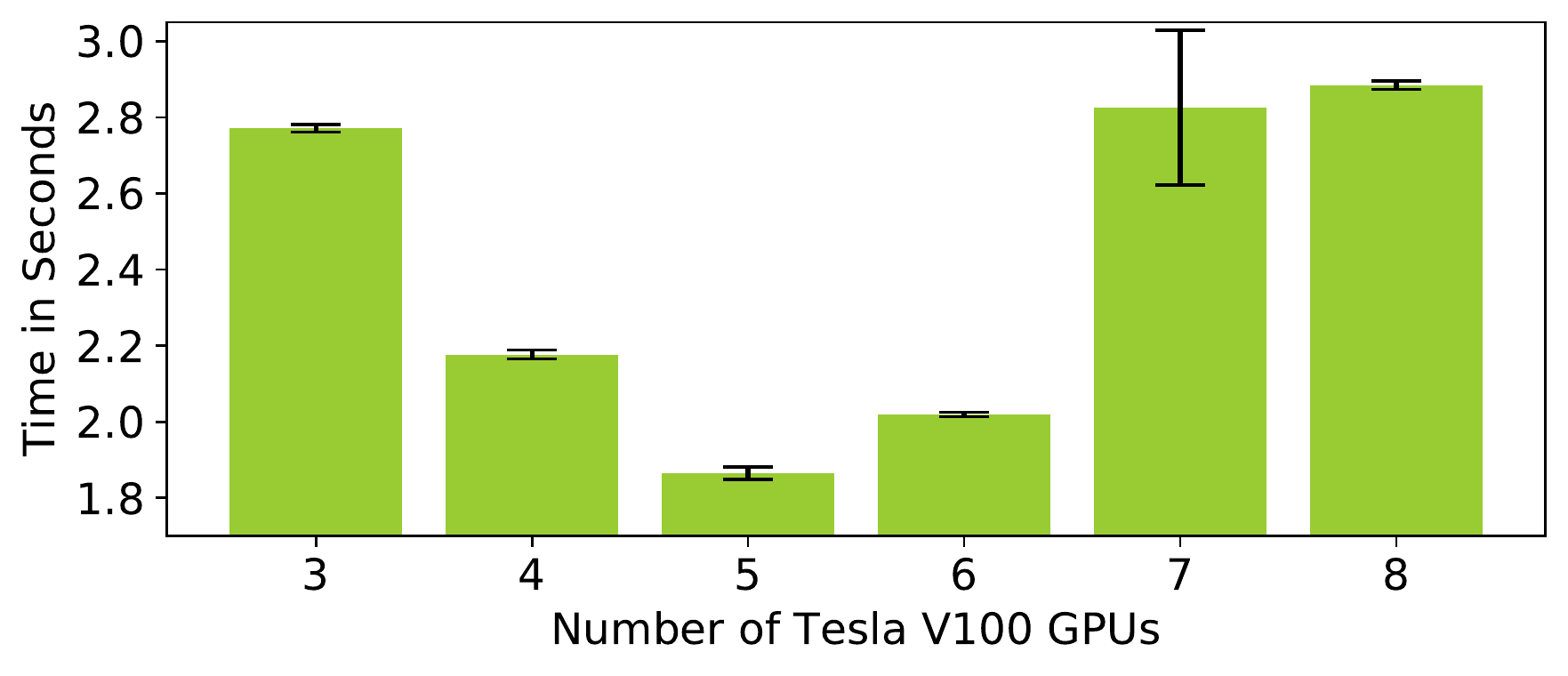}}
	\caption{Required time to solve the voxel model ``Duke'' with a $1\times 1\times 1\;\mathrm{mm^3}$ resolution and $69.8$ million DOFs on Tesla V100.}
	\label{v100_1mm}
\end{figure}
As a benchmark to test the scaling of the preconditioner and solver, eight Tesla V100 cards conjointly offer sufficient memory to increase the resolution of the human voxel model to beyond ICNIRP recommendations to $1\times1\times1\;\mathrm{mm^3}$ which is equivalent to $69.8$ million DOFs using the ``Duke'' model. Hence, the required memory should increase by about a factor of eight as well as the time to solve the Poisson system of linear equations. The results are found in Figure \ref{v100_1mm}.

The required memory is about $80$ GB and therefore at least three Tesla V100 GPUs need to be used. The best solution performance for this resolution finer than the ICNIRP recommended $2\;\mathrm{mm}$ voxel length is achieved by using five GPUs with $1.86\;\mathrm{s}$ time for a solution, which is about $8.2$ times slower than the best performance with $8.9$ million DOFs. The preconditioner and solver scale according to their near optimal asymptotical complexity in the required memory and solution time.

\subsection{Optimum Solution}
In each of the ``required time'' Figures \ref{K20m_2mm_memory}, \ref{v100_2mm}, \ref{v100_1mm} exists one bar with a large standard deviation. In Figure \ref{v100_1mm}, the bar at seven Tesla V100 cards has a large standard deviation. This inconsistency is reproducible by repeating the benchmark test of solving the system of equations. The time for solving the system is quantized and either about $2.7\;\mathrm{s}$ or $3\;\mathrm{s}$ while the memory occupation of the GPUs is identical. This effect is not understood so far.

However, it does not affect the optimum solution time of the Poisson system of linear equations for the $2\times 2\times 2\;\mathrm{mm^3}$ voxel model. The optimum solution is summarized in Table \ref{summary_req_time} for each GPU type and number of those GPUs. The shortest time to solve the system of linear equations with $8.9$ million DOFs is less than a quarter of a second using six Tesla V100 GPUs. 
\begin{table}
	\centering
	\caption{Best configuration for the tested GPUs to solve the Poisson system of linear equations for the voxel model ``Duke'' with a  $2\times 2\times 2\;\mathrm{mm^3}$ resolution. The Tesla K80 refers to a single card equipped with a dual GPU.}
	\label{summary_req_time}
	\begin{tabular}{|l|c|c|c|}
		\hline
		\textbf{GPU type}& \textbf{Tesla K20m} & \textbf{Tesla K80} & \textbf{Tesla V100} \\\hline
		\textbf{Number of GPUs}& $4$ & $2$ & $6$ \\\hline
		\textbf{Required time} & $1.287\; \mathrm{s}$ & $1.824 \; \mathrm{s}$ & $0.229 \; \mathrm{s}$\\\hline
	\end{tabular}
\end{table}
\begin{figure}[ht]
	\centering
	\centerline{\includegraphics[width=7.2cm]{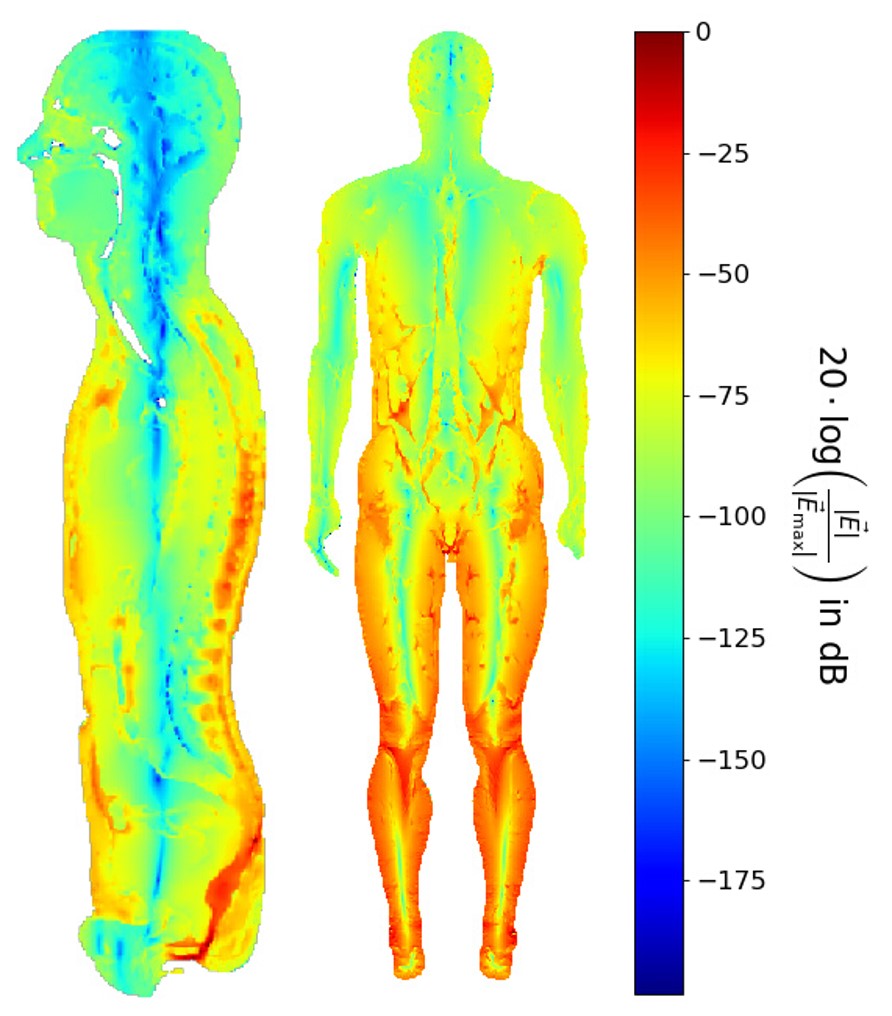}}
	\caption{Exemplary electric field strength distribution caused by a MQS field and the voxel model ``Duke'' with $2\times 2\times 2\;\mathrm{mm^3}$ resolution utilizing the Co-Simulation SPFD scheme. On the left side the median plane and on the right side the frontal plane of the human voxel model is visualized.}
	\label{E_Feld_plot}
\end{figure}
\subsection{Exemplary Electric Field Distribution}
The resulting electric field is calculated using equation (\ref{e_field}). Therefrom, the magnitude of the electric field strength $| \vec{E} |$ is calculated. Using the ICNIRP recommended voxel resolution of $2\times 2\times 2\;\mathrm{mm^3}$ yields more than $8$ million voxels. The electric field strengths are voxel averaged and the $99^{\mathrm{th}}$ percentile is taken to exclude unphysical field enhancements \cite{ICNIRP:2010}.

An example of an electric field strength distribution inside of a body is shown in Figure \ref{E_Feld_plot}.

\section{Conclusion}
 The goal of this paper was to be able to determine the exposure of a human by a MQS field caused by an inductive charging system in less than $5\;\mathrm{s}$ from magnetic field values measured in situ or simulated. Therefore, the Poisson equation of the Scalar Potential Finite Difference (SPFD) scheme had to be solved in a fraction of this time interval. The AmgX library provided solvers and preconditioners to quickly solve the Poisson system of linear equations. Even GPUs released more than half a decade ago have conjointly proven to fulfill these needs. The fastest configuration, the Tesla V100, required less than quarter of a second to solve the system of equations.
 
 The solution process on a Tesla K80 was shown to be about as fast as three Tesla K20m cards. However, the configuration of choice was either a single Tesla V100 or six Tesla V100 GPUs. A single Tesla V100 card required approximately half a second which is twice the time of six Tesla V100 cards to solve the system of linear equations with the ICNIRP recommended voxel resolution of $2\;\mathrm{mm}$ resulting into $8.9$ million DOFs using the ``Duke'' model. But the biggest advantage of the Tesla V100 compared to the Tesla K20m GPUs was the faster intercommunication speed between host and device.

\section{Acknowledgments}
This  work  was  supported  by  the  Deutsche  Forschungsgemeinschaft under grant no. CL143/14-1 and by the Bundesministerium f\"ur Wirtschaft und Energie for the TALAKO project.


\end{document}